\documentclass[useAMS,usenatbib]{mn2e}

\usepackage{graphicx}
\usepackage{txfonts}
\usepackage{rotating}
\usepackage{float}
\usepackage{multirow}
\def\lesssim{\mathrel{\hbox{\rlap{\hbox{\lower4pt\hbox{$\sim$}}}\hbox{$<$}}}}
\def\gtrsim{\mathrel{\hbox{\rlap{\hbox{\lower4pt\hbox{$\sim$}}}\hbox{$>$}}}}

\title[The Deep SPIRE HerMES Survey: Spectral Energy Distributions and
their Astrophysical Indications at High Redshift]{The Deep SPIRE HerMES Survey: Spectral Energy Distributions and
their Astrophysical Indications at High Redshift\footnotemark}
\author[D.~Brisbin et al.]
{\parbox{\textwidth}{\raggedright D.~Brisbin,$^{1}$\thanks{E-mail: \texttt{brisbind@astro.cornell.edu}}
M.~Harwit,$^{2}$
B.~Altieri,$^{3}$
A.~Amblard,$^{4}$
V.~Arumugam,$^{5}$
H.~Aussel,$^{6}$
T.~Babbedge,$^{7}$
A.~Blain,$^{8}$
J.~Bock,$^{8,9}$
A.~Boselli,$^{10}$
V.~Buat,$^{10}$
N.~Castro-Rodr{\'\i}guez,$^{11,12}$
A.~Cava,$^{11,12}$
P.~Chanial,$^{6}$
D.L.~Clements,$^{7}$
A.~Conley,$^{13}$
L.~Conversi,$^{3}$
A.~Cooray,$^{4,8}$
C.D.~Dowell,$^{8,9}$
E.~Dwek,$^{14}$
S.~Eales,$^{15}$
D.~Elbaz,$^{6}$
M.~Fox,$^{7}$
A.~Franceschini,$^{16}$
W.~Gear,$^{15}$
J.~Glenn,$^{13}$
M.~Griffin,$^{15}$
M.~Halpern,$^{17}$
E.~Hatziminaoglou,$^{18}$
E.~Ibar,$^{19}$
K.~Isaak,$^{15}$
R.J.~Ivison,$^{19,5}$
G.~Lagache,$^{20}$
L.~Levenson,$^{8,9}$
Carol J.~Lonsdale,$^{21}$
N.~Lu,$^{8,22}$
S.~Madden,$^{6}$
B.~Maffei,$^{23}$
G.~Mainetti,$^{16}$
L.~Marchetti,$^{16}$
G.E.~Morrison,$^{24,25}$
H.T.~Nguyen,$^{9,8}$
B.~O'Halloran,$^{7}$
S.J.~Oliver,$^{26}$
A.~Omont,$^{27}$
F.N.~Owen,$^{21}$
M.~Pannella,$^{21}$
P.~Panuzzo,$^{6}$
A.~Papageorgiou,$^{15}$
C.P.~Pearson,$^{28,29}$
I.~P{\'e}rez-Fournon,$^{11,12}$
M.~Pohlen,$^{15}$
D.~Rizzo,$^{7}$
I.G.~Roseboom,$^{26}$
M.~Rowan-Robinson,$^{7}$
M.~S\'anchez Portal,$^{3}$
B.~Schulz,$^{8,22}$
N.~Seymour,$^{30}$
D.L.~Shupe,$^{8,22}$
A.J.~Smith,$^{26}$
J.A.~Stevens,$^{31}$
V.~Strazzullo,$^{21}$
M.~Symeonidis,$^{30}$
M.~Trichas,$^{32}$
K.E.~Tugwell,$^{30}$
M.~Vaccari,$^{16}$
I.~Valtchanov,$^{3}$
L.~Vigroux,$^{27}$
L.~Wang,$^{26}$
R.~Ward,$^{26}$
G.~Wright,$^{19}$
C.K.~Xu$^{8,22}$ and
M.~Zemcov$^{8,9}$}\vspace{0.4cm}\\
\parbox{\textwidth}{\raggedright $^{1}$Space Science Building, Cornell University, Ithaca, NY, 14853-6801, USA: brisbind@astro.cornell.edu\\
$^{2}$Cornell University and 511 H street, SW, Washington, DC 20024-2725, USA\\
$^{3}$Herschel Science Centre, European Space Astronomy Centre, Villanueva de la Ca\~nada, 28691 Madrid, Spain\\
$^{4}$Dept. of Physics \& Astronomy, University of California, Irvine, CA 92697, USA\\
$^{5}$Institute for Astronomy, University of Edinburgh, Royal Observatory, Blackford Hill, Edinburgh EH9 3HJ, UK\\
$^{6}$Laboratoire AIM-Paris-Saclay, CEA/DSM/Irfu - CNRS - Universit\'e Paris Diderot, CE-Saclay, pt courrier 131, F-91191 Gif-sur-Yvette, France\\
$^{7}$Astrophysics Group, Imperial College London, Blackett Laboratory, Prince Consort Road, London SW7 2AZ, UK\\
$^{8}$California Institute of Technology, 1200 E. California Blvd., Pasadena, CA 91125, USA\\
$^{9}$Jet Propulsion Laboratory, 4800 Oak Grove Drive, Pasadena, CA 91109, USA\\
$^{10}$Laboratoire d'Astrophysique de Marseille, OAMP, Universit\'e Aix-marseille, CNRS, 38 rue Fr\'ed\'eric Joliot-Curie, 13388 Marseille cedex 13, France\\
$^{11}$Instituto de Astrof{\'\i}sica de Canarias (IAC), E-38200 La Laguna, Tenerife, Spain\\
$^{12}$Departamento de Astrof{\'\i}sica, Universidad de La Laguna (ULL), E-38205 La Laguna, Tenerife, Spain\\
$^{13}$Dept. of Astrophysical and Planetary Sciences, CASA 389-UCB, University of Colorado, Boulder, CO 80309, USA\\
$^{14}$Observational  Cosmology Lab, Code 665, NASA Goddard Space Flight  Center, Greenbelt, MD 20771, USA\\
$^{15}$Cardiff School of Physics and Astronomy, Cardiff University, Queens Buildings, The Parade, Cardiff CF24 3AA, UK\\
$^{16}$Dipartimento di Astronomia, Universit\`{a} di Padova, vicolo Osservatorio, 3, 35122 Padova, Italy\\
$^{17}$Department of Physics \& Astronomy, University of British Columbia, 6224 Agricultural Road, Vancouver, BC V6T~1Z1, Canada\\
$^{18}$ESO, Karl-Schwarzschild-Str. 2, 85748 Garching bei M\"unchen, Germany\\
$^{19}$UK Astronomy Technology Centre, Royal Observatory, Blackford Hill, Edinburgh EH9 3HJ, UK\\
$^{20}$Institut d'Astrophysique Spatiale (IAS), b\^atiment 121, Universit\'e Paris-Sud 11 and CNRS (UMR 8617), 91405 Orsay, France\\
$^{21}$National Radio Astronomy Observatory, P.O. Box O, Socorro NM 87801, USA\\
$^{22}$Infrared Processing and Analysis Center, MS 100-22, California Institute of Technology, JPL, Pasadena, CA 91125, USA\\
$^{23}$School of Physics and Astronomy, The University of Manchester, Alan Turing Building, Oxford Road, Manchester M13 9PL, UK\\
$^{24}$Institute for Astronomy, University of Hawaii, Honolulu, HI 96822, USA\\
$^{25}$Canada-France-Hawaii Telescope, Kamuela, HI, 96743, USA\\
$^{26}$Astronomy Centre, Dept. of Physics \& Astronomy, University of Sussex, Brighton BN1 9QH, UK\\
$^{27}$Institut d'Astrophysique de Paris, UMR 7095, CNRS, UPMC Univ. Paris 06, 98bis boulevard Arago, F-75014 Paris, France\\
$^{28}$Space Science \& Technology Department, Rutherford Appleton Laboratory, Chilton, Didcot, Oxfordshire OX11 0QX, UK\\
$^{29}$Institute for Space Imaging Science, University of Lethbridge, Lethbridge, Alberta, T1K 3M4, Canada\\
$^{30}$Mullard Space Science Laboratory, University College London, Holmbury St. Mary, Dorking, Surrey RH5 6NT, UK\\
$^{31}$Centre for Astrophysics Research, University of Hertfordshire, College Lane, Hatfield, Hertfordshire AL10 9AB, UK\\
$^{32}$Harvard-Smithsonian Center for Astrophysics, 60 Garden Street, Cambridge, MA 02138, USA}}

\begin{document}

\date{Accepted 2010 June ??. Received 2010 June ??}

\pagerange{\pageref{firstpage}--\pageref{lastpage}} \pubyear{2010}

\maketitle

\label{firstpage}

\clearpage

\begin{abstract}
The Spectral and Photometric Imaging Receiver (SPIRE) on Herschel has been carrying out deep extragalactic surveys,  one of whose aims is to establish spectral energy distributions (SED)s of individual galaxies spanning the infrared/submillimeter (IR/SMM) wavelength region.  We report observations of the (IR/SMM)  emission from the Lockman North field (LN) and Great Observatories Origins Deep Survey field North (GOODS-N).  Because galaxy images in the wavelength range covered by Herschel generally represent a blend with contributions from neighboring galaxies, we present sets of galaxies in each field especially free of blending at 250, 350, and 500 $\mu$m.  We identify the cumulative emission of these galaxies and the fraction of the far infrared cosmic background radiation they contribute.  Our surveys reveal a number of highly luminous galaxies at redshift $z\lesssim3$ and a novel relationship between infrared and visible emission that shows a dependence on luminosity and redshift.
\end{abstract}

\footnotetext{hermes.sussex.ac.uk}

\begin{keywords}
IR galaxies: spectral energy distributions: galaxy luminosities
\end{keywords}

\section{Introduction}

The Herschel Space Observatory \citep{herschel} has opened wide astronomical access to the far-infrared/submillimeter (FIR/SMM) spectral range.  With the Spectral and Photometric Imaging Receiver (SPIRE) \citep{spire}, deep cosmological surveys are studying galaxies out to redshifts of order $z \sim 3$, reaching back to epochs when the Universe was only a few billion years old.  

A primary motivation for these surveys, as well as  those undertaken with the Photodetector Array Camera \& Spectrometer (PACS) aboard Herschel, \citep{pacs} is to gain improved spectral energy distributions (SEDs) of astronomical sources.  With an appropriate redshift, integration of the flux densities demarcated by the SEDs permits derivation of rest-frame luminosities, star-formation rates, and other physical properties of galaxies. 

Among the first observations undertaken by SPIRE in the Herschel Multi-tiered Extragalactic Survey (HerMES\footnotemark[2]) project \citep{hermes} have been surveys of galaxies in GOODS-N and Northern portions of the Lockman Hole (LN) field [see \citet{o2} for a description of these early observations.]  Source confusion, as defined and discussed in detail by \citet{tt}, results in blending of far-infrared sources and complicates the analysis of survey data.  In light of the large degree of source blending expected at SPIRE wavelengths, novel options for source extraction have been pursued [\citep{i1}, \citep{s2}, \citep{b2}].   Rather than looking  for sources based on SPIRE intensity maps alone or relying on traditional source detection and extraction techniques for the SPIRE data, which are heavily affected by confusion, \citet{i1} measure the SPIRE flux at the position of known 24 $\mu$m sources using a linear inversion technique to account for source blending. The rationale for this is provided by the results of the Balloon-borne Large Aperture Submillimetre Telescope (BLAST) extragalactic survey \citep{m1}, which showed that the 24 $\mu$m and the FIR flux densities are at least statistically correlated.

\section{Primary Aims}

The aims of this paper are twofold; our primary aim is to derive spectral energy distributions for distant galaxies observed by SPIRE.  Before this can be achieved, however, a robust way of identifying sources least affected by confusion and blending must be devised.

The GOODS-N catalogue of \citet{i1} provides a cross-identification (XID) of FIR/SMM flux density at 250, 350 and 500 $\mu$m with 1951 possible 24 $\mu$m counterparts having minimum flux densities of 20 $\mu$Jy.  Many of the identified 24 $\mu$m galaxies are further cross-identified with ultraviolet, optical, near-infrared (NIR) and radio counterparts. The survey covered a  $12.3 \times 18.6 $ arc minute strip, corresponding to $\sim$230 arcmin$^2$

\begin{figure}
	\centering
		\includegraphics[height=.23\textheight]{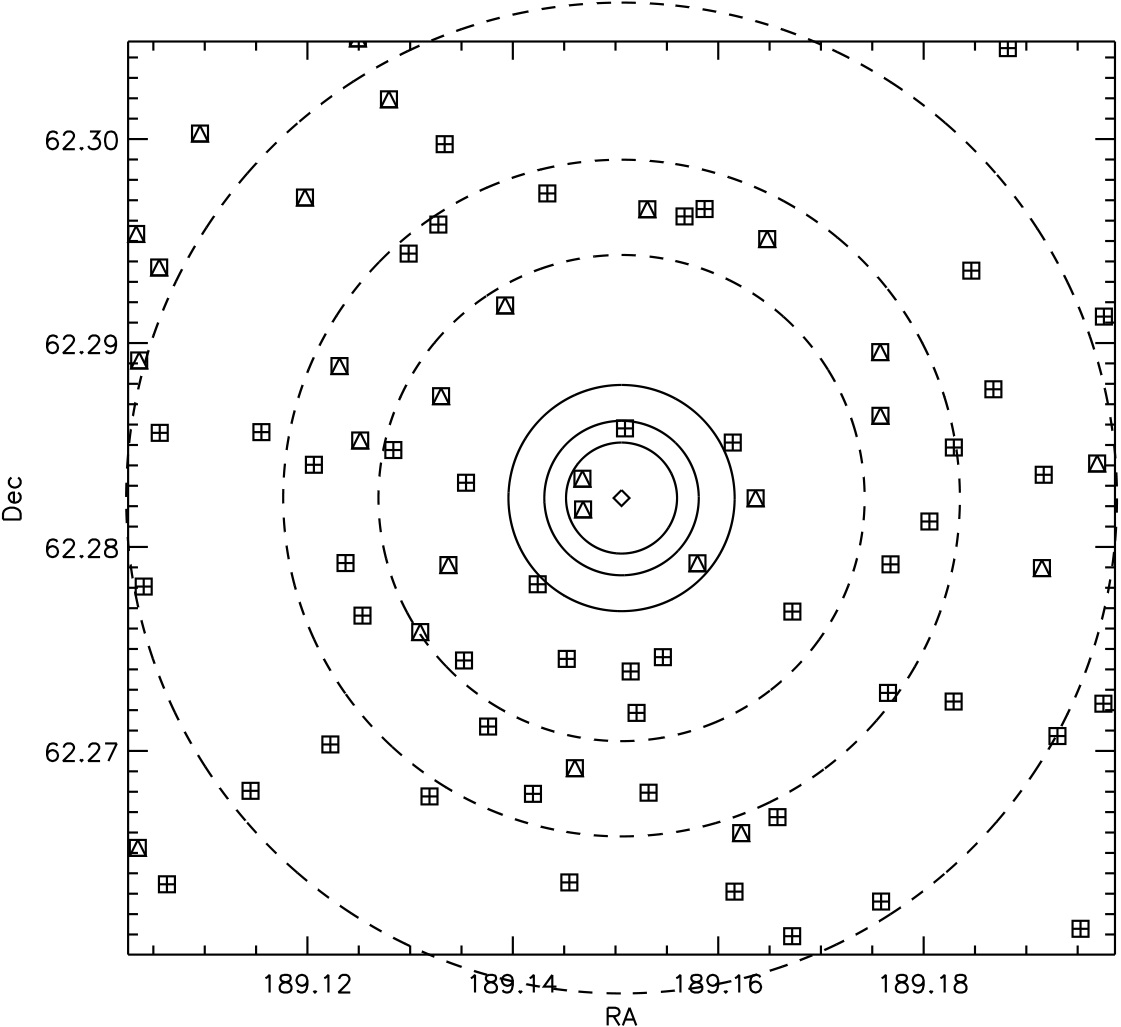}
	\caption{A GOODS-N SPIRE and 24 $\mu$m source 
	with equatorial coordinates (J2000.0) 189.1506 / 62.2824 shown as a diamond with nearby 24 $\mu$m sources.  Known ambient 24 $\mu$m sources are shown as squares with nested symbols.  Plus signs indicate sources with known redshifts and triangles indicate sources with unknown redshift.   The spatial distribution of 24 $\mu$m sources is based on the XID information in \citet{i1}.  Solid circles represent the full-width-half-maxima (FWHM) of the Airy profile PRF at the three SPIRE wavelengths and dashed circles represent the second Airy minimum.  Coordinates are in J2000.}
	\label{fig:crowdedbeam}
\end{figure}

The SPIRE beam diameters at full-width-half maximum (FWHM) respectively measure 18.1, 25.2 and 36.9 arcsec at 250, 350 and 500  $\mu$m.   For present purposes, we take the beams to be close to circular; their ellipticity varies from pixel to pixel, but is approximately $1.08 \pm 0.05$, the longer direction lying in the spacecraft horizontal direction, parallel to the ecliptic plane \citep{sc1}. The beam at 500 $\mu$m thus has an area $\sim$0.3 arcmin$^2$.  With 1951 possible 24 $\mu$m sources, we can expect a typical 500 $\mu$m beam to contain 2.5 possible sources.  At 250 $\mu$m the crowding is a factor of 4 less severe, but still appreciable.  An example of the crowded source distribution is seen in Figure \ref{fig:crowdedbeam} where the SPIRE beam outlines are overlaid on a patch of the GOODS-N field centered on a 24 $\mu$m source.

In the Northern Lockman region, a 40.1' $\times$ 36.2' segment of the sky yielded $6316$ possible 24 $\mu$m counterparts with minimum flux densities of 50 $\mu$Jy in an area subtending $\sim$1500 arcmin$^2$ --- again corresponding to more than 1 potential 500 $\mu$m source per beam.  Identification of sources least affected by confusion and blending is therefore important.  These sources have also been cross-identified at multiple wavelengths and assigned photometric redshifts as detailed in \citet{st}.


\section{Data}
\footnotetext[1]{Herschel is an ESA space observatory with science instruments provided by European-led Principal Investigator consortia and with important participation from NASA.}
\footnotetext[2]{hermes.sussex.ac.uk}

Using the existing high spatial resolution Spitzer 24 $\mu$m data and the known far-infrared instrumental point response function (PRF) as  inputs, \citet{i1} determined best-fit 250, 350, and 500 $\mu$m fluxes by a procedure they detail in their paper.  In each SPIRE wavelength band, their tabulated cross-identifications provide both their best estimate of the flux density $F_{\nu}$ and the flux density $PRF_{\nu}$ in a PRF-convolved map centered on the position of an associated 24 $\mu$m source.  They make no assumptions about a proportionality between 24 $\mu$m and FIR flux densities, but assume that SPIRE sources will only be detected at positions of 24 $\mu$m sources.  
For each source, ancillary data at other wavelengths are included, as well as several flags that can be used to identify degenerate cases.   Of particular importance is the redshift of the associated 24 $\mu$m source, which enables derivation of rest-frame SEDs and thus source luminosities.  Wherever we refer to $F_{\nu}$ and $PRF_{\nu}$ in the remainder of this paper these quantities are to be thought of as those defined by \citet{i1}.


The flux density in a PRF-convolved region centered on the position of the 24 $\mu$m galaxy associated with each far-infrared source represents the system response not only to the flux density attributed to this source (referred to as $F_{\nu}\bigr|_\lambda$) but also to contributions from nearby  sources.  It thus provides a measure of  blending characterizing each source.  We define a {\it purity index} $\Pi_{\lambda}$ for each source as the ratio 
\begin{equation}
\Pi_{\lambda} = F_{\nu} /PRF_{\nu}\bigr|_{\lambda}\ ,
\end{equation}
where $\lambda$ specifies the wavelength band, 250, 350, or
500 $\mu$m, and $PRF_{\nu}\bigr|_\lambda$, as supplied in the XID catalogs by Roseboom et al. 2010  is the
PRF-smoothed flux density at the position of the source.\footnotemark[3]
A high value of $\Pi_{\lambda}$ indicates low confusion and blending in wavelength band $\lambda$; a low value indicates high blending. The fractional contributions by ambient sources to the PRF-convolved flux density within a PRF is simply $(1 - \Pi_{\lambda})$.  In principle, the purity index must assume values $0\lesssim\Pi_{\lambda}\lesssim 1$. In practice, however, the value $F_{\nu}$ has been calculated {\it on top of} a locally determined background, whereas the $PRF_{\nu}$ values do not take a local background variation into account.  This can result in $\Pi_{\lambda}>1$ when a negative local background or ``baseline" has been used.

\footnotetext[3]{Our $PRF_{\nu}\mid_{\lambda}$ corresponds to the quantity ${\bf d}$ in
equation (2) of Roseboom et al. 2010, convolved with the point response
function centered on the primary source whose flux is $F_{\nu}\mid_\lambda$.
The entries in the XID tables list our $F_{\nu}\mid_\lambda$ as
$F(\lambda)$, and our $PRF_{\nu}\mid_\lambda$ flux density as
$PRF(\lambda)$.}
                                                                
\section[]{The Statistics of Purity Indices in GOODS-N and Lockman North}

We have found it useful to identify sources whose purity indices respectively exceed $\Pi_{\lambda}$ measures of 0.7, 0.5, and 0.3 at 250, 350, and 500 $\mu$m.  We say that a source is {\it secure} at each wavelength if it meets this criterion.  If the source is found to be secure in all three wavelength bands, we call it {\it triply secure}.  High-purity sources are of special interest because their isolated nature makes them less susceptible to blending by neighbors.  It is these sources which will be especially useful for follow up studies with other instruments and also may provide confirmation of the deblending approach used.  It should be noted, however, that a flux density estimate from a highly pure source might still be inaccurate if there is significant contribution from an infrared source that is not observed at 24 $\mu$m.  Furthermore, sources with low purity do have well-defined deblending solutions and hence well-characterized flux densities and uncertainties from \citet{i1}.   In crowded fields, the true flux density is very likely described by this characterization, although the margins of uncertainty tend to be large.  Our choice of purity criteria is somewhat subjective but offers a reasonable compromise for extracting relatively reliable SEDs despite source blending.  Understandably, these criteria may be expected to vary depending on the type of information an astronomer expects to extract from the survey data.

When a reliable redshift is available, the single most important quantity
that can be determined from an SED is source luminosity.  With this, one can
begin discussing the luminosity distribution at specific redshifts, as well
as luminosity evolution as a function of redshift, particularly among
ultraluminous galaxies that emit the dominant fraction of their energy in
the infrared. However, to obtain a reliable SED and thus a reliable luminosity,
we require sources whose flux densities are well determined at all three SPIRE
wavelengths in order to optimally constrain the flux density defining the
broad wavelength region around peak emission.

To explain the consequences of our choice of purity criteria in this
context, we may consider a toy model which, as pointed out in Section 1,
will respectively exhibit an average number of sources $n_{\lambda} \sim 2.5$, 1.225, and
0.625 per GOODS-N beam, at 500, 350, and 250 $\mu$m.  Let us
inject an additional source into such a beam and call it the primary source.
If all the sources involved are equally bright, on average, i.e., make equal
contributions to the PRF-smoothed flux density, the purity of the primary
source will be $\Pi_{\lambda} = (1 + n_{\lambda})^{-1}$, i.e.  
0.29, 0.45, and 0.62, respectively  at 500, 350 and 250 $\mu$m.  Half the sources in each waveband will have purities higher than these purity cuts, and half lower.

Turning now to our preferred adoption of purity cuts of 0.3, 0.5 and 0.7 at 500, 350, and 250 $\mu$m, we see that they assure two properties: (i) that they yield sources whose purities are above average at all three wavelengths, and (ii) that the fraction of sources with purity above the cut is roughly comparable at all three wavelengths --- a balance, which is important to assure a well-defined SED.  Table 1, described below, confirms these traits for the GOODS-N sample.  It shows that a fraction $f_{\lambda} = 0.23$ of the  sources has purity exceeding 0.7 at 250 $\mu$m, a fraction 0.32 exceeding purity 0.5 at 350 $\mu$m, and a fraction 0.36 exceeding purity 0.3 at 500 $\mu$m.  These fractions cluster around a value of 0.3, thus lending roughly equal weight to the flux density in each waveband in the determination of the SED.

In Table 1, we list the fraction of sources in GOODS-N and LN whose purity indices lie above certain cuts.  We permit these indices to slightly exceed a value of one, with a cut-off of $\Pi = 1.1$ in GOODS-N and $\Pi = 1.2$ in LN.  These relaxed upper limits are designed to allow the inclusion of detections with a significant local background which otherwise have all the earmarks of being secure.

In the GOODS-N region the XID catalog lists 183 sources observed at all
three wavelengths and with known redshift, 16 of which are triply secure. If we remove the upper
limit $\Pi = 1.1$, the number of triply secure sources rises to 59. In LN
there are 633 sources with detections at all five wavelengths with known redshift, 165 of which
are triply secure; this number increases to 287 if we remove the upper
limits on purity.  Although the numbers of these galaxies are quite modest,
they nevertheless yield informative statistics on the luminosities and
luminosity distributions of galaxies observed out to redshifts $z\sim 3$.
These will be discussed in Section 6.

\begin{table}
\caption{Fraction of detected SPIRE sources in GOODS-N and LN with a $\Pi_{\lambda}$ value within the range indicated in the top row.  The columns marked ``detections" denote the total number of SPIRE sources detected at a given wavelength in the current HerMES survey.}
\centering
\begin{tabular}{|l|l|l|l|l|l|l|l|l|l|}
\hline
\multicolumn{6}{|c|}{GOODS-N} \\ 
1.1$>\Pi>$ & 0.9 & 0.7 & 0.5 & 0.3 &   detections  \\ \hline
250 $\mu$m & 0.106 & 0.231 & 0.348 & 0.493 &  1032 \\ 
350 $\mu$m & 0.069 & 0.199 &  0.316 & 0.451 &  697 \\ 
500 $\mu$m & 0.061 & 0.141 &  0.227 & 0.362 &  475 \\ \hline \hline
\multicolumn{6}{|c|}{LN} \\
1.2$>\Pi>$ & 0.9 & 0.7 & 0.5 & 0.3 &   detections  \\ \hline
250 $\mu$m & 0.275 & 0.435 & 0.579 & 0.703 &  4646 \\
350 $\mu$m & 0.184 & 0.343 &  0.500 & 0.670 &  2968 \\
500 $\mu$m & 0.144 & 0.281 &  0.419 & 0.570 &  2127 \\
\end{tabular}
\label{tab:PIstats2}
\end{table}

The larger aperture of the Herschel telescope and the higher spatial resolution  this enables have permitted the SPIRE surveys to reach depths beyond those attained by BLAST.  Nevertheless, \citet{m1} succeeded in acquiring reliable measurements of stacked source flux densities at comparable wavelengths.  Their results indicate surface brightnesses of 8.60 $\pm$ 0.59, 4.93 $\pm$ 0.34, and 2.27 $\pm$ 0.20 nW m$^{-2}$ sr$^{-1}$ at 250, 350, and 500 $\mu$m respectively.  These stacked source flux densities represent the major component of the cosmic infrared background (CIB) measured by the Cosmic Background Explorer's Far-Infrared Absolute Spectrometer (FIRAS) to be 10.4 $\pm$ 2.3, 5.4 $\pm$ 1.6, and 2.4 $\pm$ 0.6 nW m$^{-2}$ sr$^{-1}$ at 250, 350, and 500 $\mu$m respectively \citep{f1}.  To investigate the extent to which this background is resolved with Herschel, we summed the estimated flux densities for our secure SPIRE sources in the deepest field (GOODS-N) and then attributed this flux density to the entire survey region of 230 arcmin$^2$.  Using the flux densities for sources contained in the XID catalog of  \citet{i1}, our cumulative surface brightness for GOODS-N at 250, 350, and 500 $\mu$m is 1.49, 0.70, and 0.41 nW m$^{-2}$ sr$^{-1}$ or 14\%, 13\%, and  17\% of the estimated CIB.  At 250 and 350 $\mu$m these values are within 1$\sigma$ of those corrected for blending and incompleteness by \citet{o2}.

\section[]{Spectral Energy Distribution of the Secure Sources}

\begin{figure}
	\centering
		\includegraphics[width=.5\textwidth]{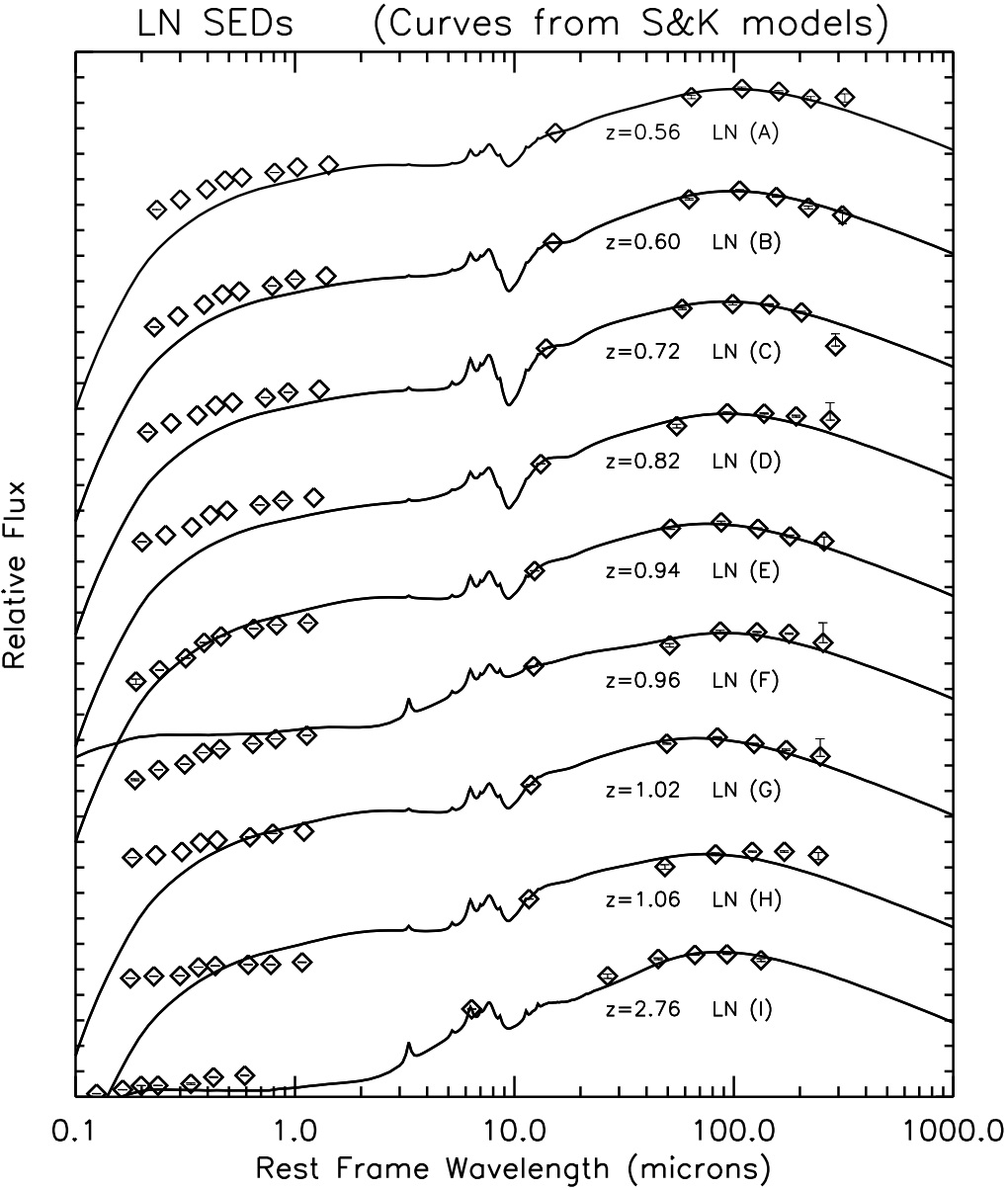}

	\caption{Lockman N galaxy SEDs plotted with arbitrary flux density offsets.  These galaxies all met or exceeded our selection criteria of being securely identified at all three wavelengths, having a high purity index, (30\%, 50\%, and 70\% pure at 500, 350, and 250 $\mu$m respectively), having known PACS detections at 100 and 170$\mu$m, and having known redshifts, $z>0.5$.  Each tick mark on the y-axis indicates a change by a factor of 10.  Observed flux densities are indicated by diamonds.  For most observations the error bars are smaller than the diamond; however for LN sources C, D, F, and G the 500 $\mu$m measurement is compatible with zero, so we have plotted only the upper error bar.  The solid line is an S\&K model fit to the 24 $\mu$m through 500 $\mu$m error-weighted observations.  We weighted the 24 $\mu$m observations only a quarter as heavily as the longer wavelengths  which play a dominant role in determining starburst luminosities.  Observational data shortward of 24 $\mu$m is plotted for reference but not used in fitting.  Source $\Pi$s are shown in Table \ref{tab:bigLN}.  As with many of the examples quoted by S\&K, the visual component of the SED often needs to be fitted by hand because it bears little relation to the starburst characteristics responsible for the mid- and far-infrared flux densities.  The extent to which visible stars may or may not contribute to the SED is determined in part by the degree to which an older population of stars is obscured by dust without significantly contributing to its heating or by massive young stars whose shrouding by dust has gradually declined.  Shrouding by dust may explain the significant drop in optical luminosities exhibited by some of the sources at the shortest wavelengths.}
	\label{fig:stackedLNsed}
\end{figure}

\begin{figure}
	\centering
		\includegraphics[width=.5\textwidth]{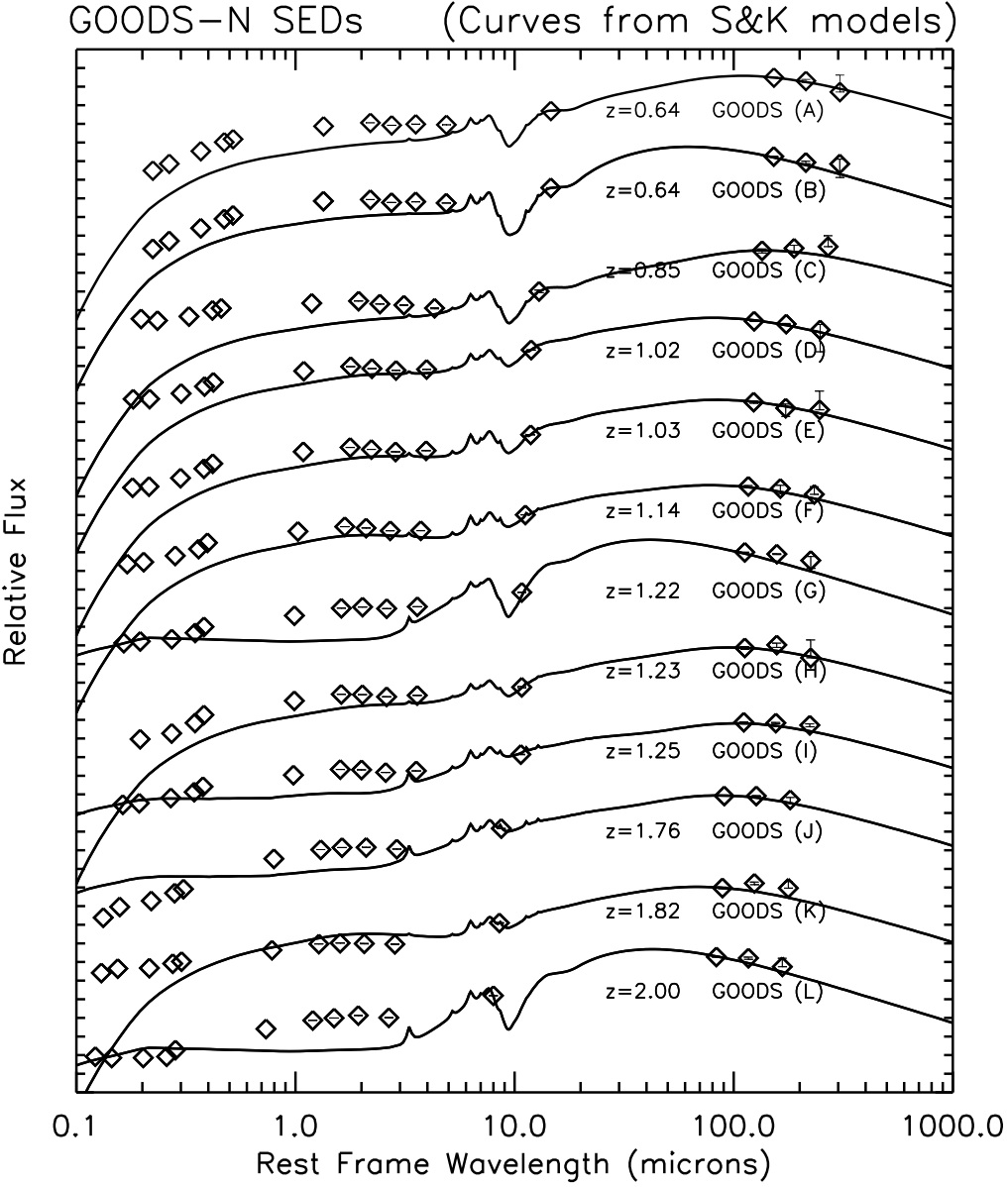}
	
	\caption{ GOODS-N galaxy SEDs plotted as in Figure \ref{fig:stackedLNsed}.  The 500 $\mu$m measurements for GOODS sources A, C, E, F, H, K, and L are compatible with zero, so here we plot only the upper error bar, barely visible within the diamond symbol.  As in Figure \ref{fig:stackedLNsed}, we weighted the 24 $\mu$m data a quarter as heavily as the longer wavelengths.  Observational data shortward of 24 $\mu$m is plotted for reference but not used in fitting.}
	\label{fig:stackedGOODSNsed}
\end{figure}

Figures \ref{fig:stackedLNsed} and \ref{fig:stackedGOODSNsed} exhibit SEDs for triply secure sources in LN and GOODS-N.  For LN sources, we have set an additional criterion for inclusion, namely that they have observed flux densities also at PACS wavelengths of 100 and 170 $\mu$m.  Along with examining the observed SED, we show a fit using starburst models developed by \citet{s1} (S\&K).  These models are based on a nuclear concentration of massive young stars embedded in a matrix of gas and dust referred to as ``hot spots".   S\&K use a five parameter SED fit which incorporates a variable nuclear bulge size with old and new stellar components as well as the effects of dust. They provide their models in the form of a library of 7000 SEDs available as text files online.\footnotemark[4]
By using their models, we are able to not only find realistic intrinsic luminosities, but also star formation rates (SFR) for highly luminous sources at high redshift for which the Kennicut infrared - SFR relations apply \citep{k1}.

\footnotetext[4]{The S\&K SED library of models is available at http://www.eso.org/$\sim$rsiebenm/sb\_models/.}

The S\&K model fits observations quite well, although some of our SEDs exhibit considerable deviations from the data at visible and near infrared wavelengths.  This is largely  due to variable shielding of starlight by dust, see section 6 below.

In Figures \ref{fig:stackedLNsed} and \ref{fig:stackedGOODSNsed}, we have focused on high redshift (z$>$0.5) sources as these are of greatest interest to star formation history. 
Tables \ref{tab:bigLN} and \ref{tab:bigGOODSN} show the luminosities and star-formation rates of galaxies whose SEDs are shown in Figures 2 and 3.  The models appear capable of correctly representing the galaxies we observe, most of which are highly luminous and likely to be starbursts.  Like most current models, however, vastly different dust masses and nuclear sizes are able to yield similar SEDs, as witnessed by the large differences in dust masses assigned to some galaxies listed in Tables \ref{tab:bigLN} and \ref{tab:bigGOODSN} that have nearly identical redshifts and luminosities.  This is not surprising because the models are only required to provide sufficient dust to convert most of the visible and ultraviolet radiation produced in the starburst into infrared radiation at the observed temperature. If this criterion is satisfied, the models produce roughly correct SEDs.

While these figures and tables emphasize the relatively few triply-secure SEDs among high redshift objects, it is important to note that for many statistical trends, triple-security is not necessary.  In Figure \ref{fig:lumdist}, we show the luminosities we have determined as a function of redshift, based on the S\&K models for all sources in GOODS-N and LN that have detections of any kind (secure or not) at all three SPIRE bands.  While some of the flux densities, especially at 500 $\mu$m, are quite uncertain (see Figures \ref{fig:stackedLNsed} and \ref{fig:stackedGOODSNsed}), this will not greatly affect the overall luminosity distribution.  We estimate the source luminosity uncertainty to be $\sim$20\%.  The minimum detectable luminosities follow the expected trend of increasing with the square of luminosity distance (which we determined based on a $\Lambda=0.7$, H$_0=70$ km s$^{-1}$ Mpc$^{-1}$ cosmology).

Throughout this paper we refer to multiple luminosities, namely total bolometric luminosity and infrared luminosity integrated over 8 - 1000 $\mu$m [as used in star formation estimates by \citet{k1}].  For our purposes, the difference between the two is small as the majority of a luminous star-forming galaxy's energy is emitted in the infrared.  In galaxies with $L_{IR} \gtrsim 10^{11.5} L_\odot$, \citet{buat} find that $\sim$95\% of the total star formation rate is accounted for by $L_{IR}$.  Nonetheless, we explicitly differentiate between the two when the distinction is significant.

\section{The Nature of the Ultraluminous Sources} 

As Figure 4 attests, our surveys reveal a number of highly luminous sources, mainly at redshifts between $z=2.5$ and 3.2.  Discussing all of these in detail is beyond the scope of the present paper, if for no other reason than that information on these sources is still quite modest. Nevertheless, we here discuss four of these sources to show that their luminosities appear to be intrinsic. Neither gravitational lensing nor blending from neighboring sources appear to contribute significantly to the observed luminosities. This may not be true of all of the luminous SPIRE sources, but it appears to be so for at least those about which we have the most information right now.  A number of other uncertainties also warrant comment. 

(i) For most of the sources observed in LN only photometric redshifts are available.  For the more luminous sources shown, estimated redshift errors range from $\sim 6$ to 17\% in the catalogs of \citet{i1}.  Because calculated luminosities are proportional to $(1+z)^2$, a 17\% error at $z = 2.5$ can lead to a luminosity error in the range of $\sim 23 - 26\%$.  For sources with established spectroscopic redshifts, the uncertainty is far smaller and can be neglected.

(ii) A second area of concern is discrimination between intrinsically luminous sources and sources that 
may be lensed by foreground galaxies to merely appear luminous.  For some purposes, e.g. accounting for sources contributing to the diffuse background, this distinction may not be of primary importance; but for charting the luminosities of distant sources their intrinsic luminosities need to be determined.  

(iii) Because our deep surveys are highly sensitive, they reveal a large number of faint sources.  This leads to potential mis-identification of sources.  It can also lead to blending.   But seven of the eleven ultraluminous sources in Figure 5 and Table 4 are triply secure; the remaining four are doubly secure within observational uncertainties.  Severe blending is thus unlikely.

(iv)  For some of the sources, the available data points straddle but do not directly constrain the region where the SED reaches a maximum (see Figure 5), so that our SED-derived luminosities could be over-estimated.  

At the present stage of data reduction we are not yet in a position to account for all these uncertainties.  However, for the cited sources in GOODS-N and for a few of the sources in LN reliable spectroscopic redshifts are available.  Among four of these we were able, below, to search for potential lensing, assess a degree of blending, and justify our confidence in their derived luminosities in some detail.  These sources are referred to by letters corresponding to their designations in Table \ref{tab:bigluminous}.  Note that in Table \ref{tab:bigluminous} we give bolometric luminosities whereas below we quote infrared luminosities.

The LN source (j) at equatorial position (J2000.0) 161.554052 / 58.788592,  is characterized in the Sloan Digital Sky Survey (SDSS) as a well-isolated circular source identified as a quasi-stellar object (QSO). The visible spectrum leaves no doubt about the redshift $z =3.037$ determined by the strong Ly-$\alpha$ line.  The visible continuum flux in this spectrum  is $\sim 5\times 10^{-17}$ erg cm$^{-2} $ s$^{-1}$ \AA$^{-1}$, equivalent to $\sim 4\times 10^{-5}$ Jy, stretching down to $\sim 1,000$ \AA, for a total flux of roughly $1.2\times 10^{-15}$ W m$^{-2}$, which is comparable to the infrared flux observed from the source.  The IPAC extragalactic data base lists the optical source as brighter, by more than a magnitude, than any other source within a radius of an arc minute. \citet{t1} list a 0.5-2 keV X-ray flux of $38.4\times 10^{-18}$ W m$^{-2}$, and a 2-8keV flux of $22.4 \times 10^{-18} $W m$^{-2}$ for the source, jointly about 20 times lower than the visible flux.  \citet{o1} detected a 20-cm continuum radio flux density of 64 $\mu$Jy from this source, within an apparent size $< 1.5$ arcsec FWHM, ruling out obvious ambient emission that might have indicated lensing. The XID catalogs of \citet{i1} show a relatively weak neighboring SPIRE source at a distance of 26 arcsec, and a source comparable in infrared brightness to LN 4241 but at a separation of $\sim 34$ arcsec.  Because of their relatively large displacement, these and other ambient galaxies are unlikely to contribute appreciably to the SPIRE flux densities assigned to LN (j) by \citet{i1}.  All this gives confidence that both the visible and the infrared flux come from the same source, that there is no lens magnification, and that we are indeed dealing with an un-lensed ultraluminous source with infrared luminosity $\sim 1.8\times 10^{13}L_{\odot}$.  The simplest explanation for these data is that we are viewing a QSO with a surrounding torus along a sightline coinciding with the torus axis.  The visible light reaches us directly along this axis; the infrared emission comes from dust heated in part by the QSO and possibly also by massive star formation. 

The GOODS-N source (c), with equatorial coordinates (J2000.0) 188.990097 / 62.17342, is cited by \citet{b1} as having  specroscopic redshift $z=3.075$. The 24 $\mu$m flux density listed in the Barger catalog is  109 $\mu$Jy. The rest-frame 2-8keV luminosity assigned to this source by \citet {t1} is $3.464 \times 10^{36}$W $\sim 9.01 \times 10^9 L_{\odot}$, but their paper provides no rest-frame 0.5-2keV luminosity. The nearest SPIRE source listed in the XID catalogs  lies at a distance of 35 arcsec, where its contribution to our primary galaxy's flux listed in the XID catalog can at most be minor.  Our primary source also displays weak X-ray fluxes, but \citet{m2} list no 20-cm source within 3.5 arcmin.  Integrating the flux densities indicated by the fitted SED in Figure \ref{fig:luminoussed}, leads to an infrared luminosity of $6.0\times 10^{12} L_{\odot}$.

One of our ultraluminous sources [LN (f)] has previously been discussed by \citet{p1}. They observed the Lockman SWIRE source at (J2000.0) 161.041521 / 58.87355 with Chandra and detected a flux of $2.7 \pm 1.1 \times 10^{-15}$ erg  cm$^{-2}$ s$^{-1}$ in the 0.3 - 8 keV range. The Spitzer 24 $\mu$m flux is 4.0 mJy, strong for a source at spectroscopic redshift 2.54, and much brighter than anything within an arcminute of its location. Because of its initial detection by Spitzer, the authors characterize the source as an infrared selected Compton-thick AGN on the basis of the rest-frame hydrogen column density, which they estimate to be $N_H \sim 3\times 10^{24}$ cm$^{-2}$ with an uncertainty envelope extending a factor of three times lower and arbitrarily higher. The infrared luminosity derived on the basis of our SPIRE and PACS observations (see Figure \ref{fig:luminoussed} and Table \ref{tab:bigluminous}) is $2.0\times 10^{13}L_{\odot}$.   It appears to be fairly well isolated in the infrared, the nearest comparably bright 250 micron source being located half an arcminute away.  

Some of these ultraluminous galaxies could be lensed but a first look has not yet revealed these in our sample.  The GOODS-N source  (d), with coordinates (J2000.0) 189.309509 / 62.20232, is cited by \citet{b1}  as having a spectroscopic redshift $z=3.157$. Again, ambient nearby sources have relatively weak SPIRE fluxes unlikely to appreciably affect the SPIRE flux attributed to our source of primary interest.   Another source only $\sim 10$ arcsec away is also noted in the NASA/IPAC Extragalactic Database (NED). This appears not to have a measured 24 $\mu$m flux and is not listed by \citet{b1}.   However, \citet{l1} have included  this source in their discussion of distant irregular galaxies. The object designated as BX 150 appears elongated roughly along a north/south direction, is $\sim 0.5$ arcsec long  with an aspect ratio roughly 2:1, and has a redshift $z= 2.28$.  At optical wavelengths, it is 1.3 magnitudes fainter than the ultraluminous infrared source and, at its rather high displacement of $\sim 10$ arcsec, it is unlikely to provide significant lensing.  With this proviso, the intrinsic infrared luminosity of GOODS (d) appears to be $\sim 1.1\times 10^{13}L_{\odot}$.

Figure \ref{fig:luminoussed} provides the SEDs of these ultraluminous sources. Our combined surveys of GOODS-N and LN cover $\sim 0.47$ square degrees, or one part in 85,000 of the sky.  Given that we observe several high-luminosity sources in the small area covered, it suggests that approximately $10^6$ sources in the infrared luminosity range $\sim 10^{13} L_{\odot}$ should be observable in the Universe, at redshifts $z=2.5$ to 3 at the current epoch.  The number of comparably luminous sources observable at lower redshifts appears to sharply decline.  

\begin{figure}
	\centering
		\includegraphics[width=.5\textwidth]{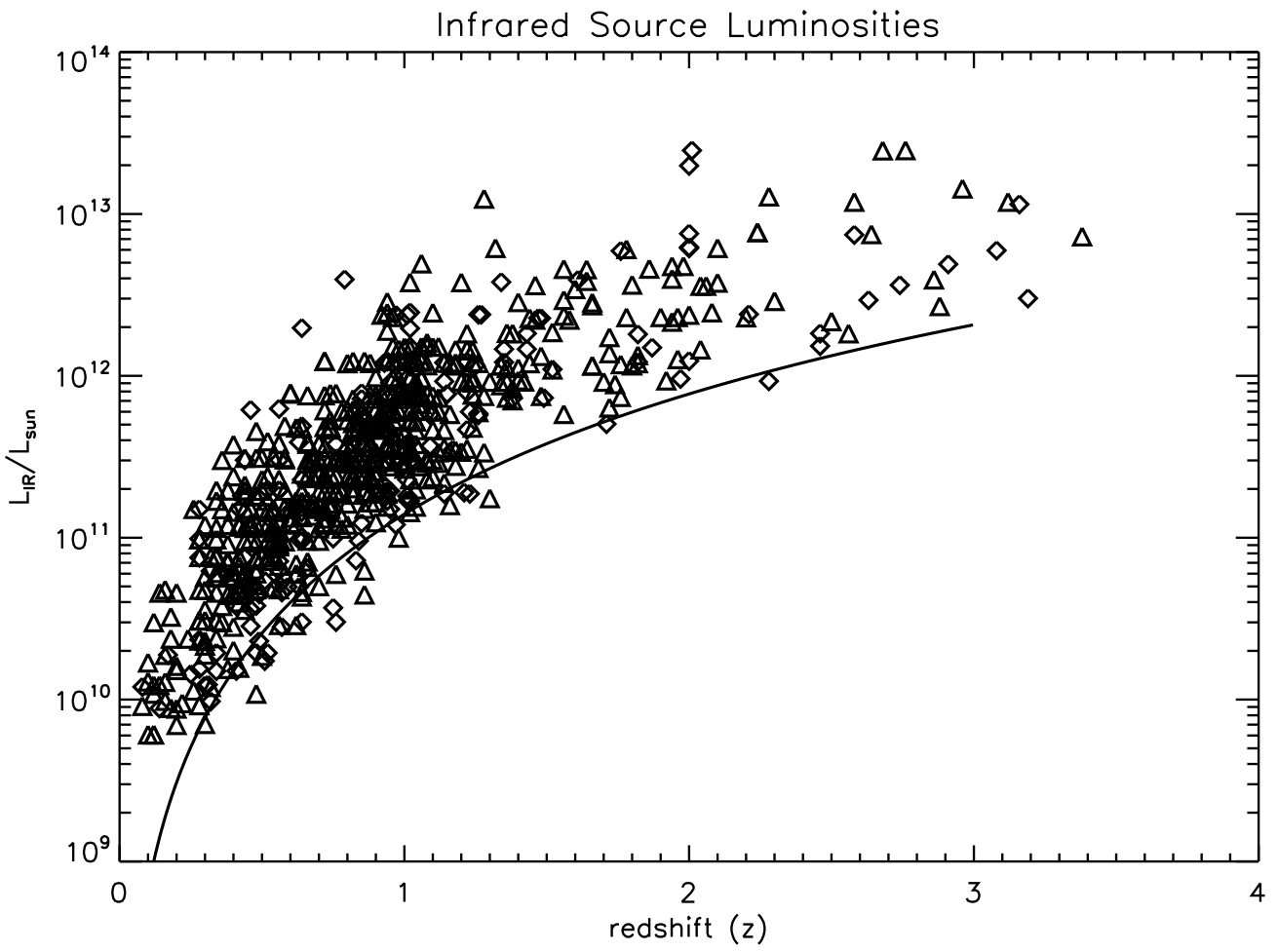}
	\caption{Infrared source luminosities (integrated over 8-1000 $\mu$m) in GOODS-N and LN plotted as a function of redshift for all sources detected at all three SPIRE wavelengths.  Diamonds indicate GOODS-N luminosities obtained from SEDs fitted by S\&K models, and triangles indicate similar luminosities for LN sources.  The solid line shows the growth of luminosity distance squared with z.  It serves as a rough lower bound to the luminosities in our selection of observed sources; the scatter of data points about the curve provides a visual impression of the uncertainties in those luminosities.}
	\label{fig:lumdist}
\end{figure}

\begin{figure}
	\centering
		\includegraphics[width=.5\textwidth]{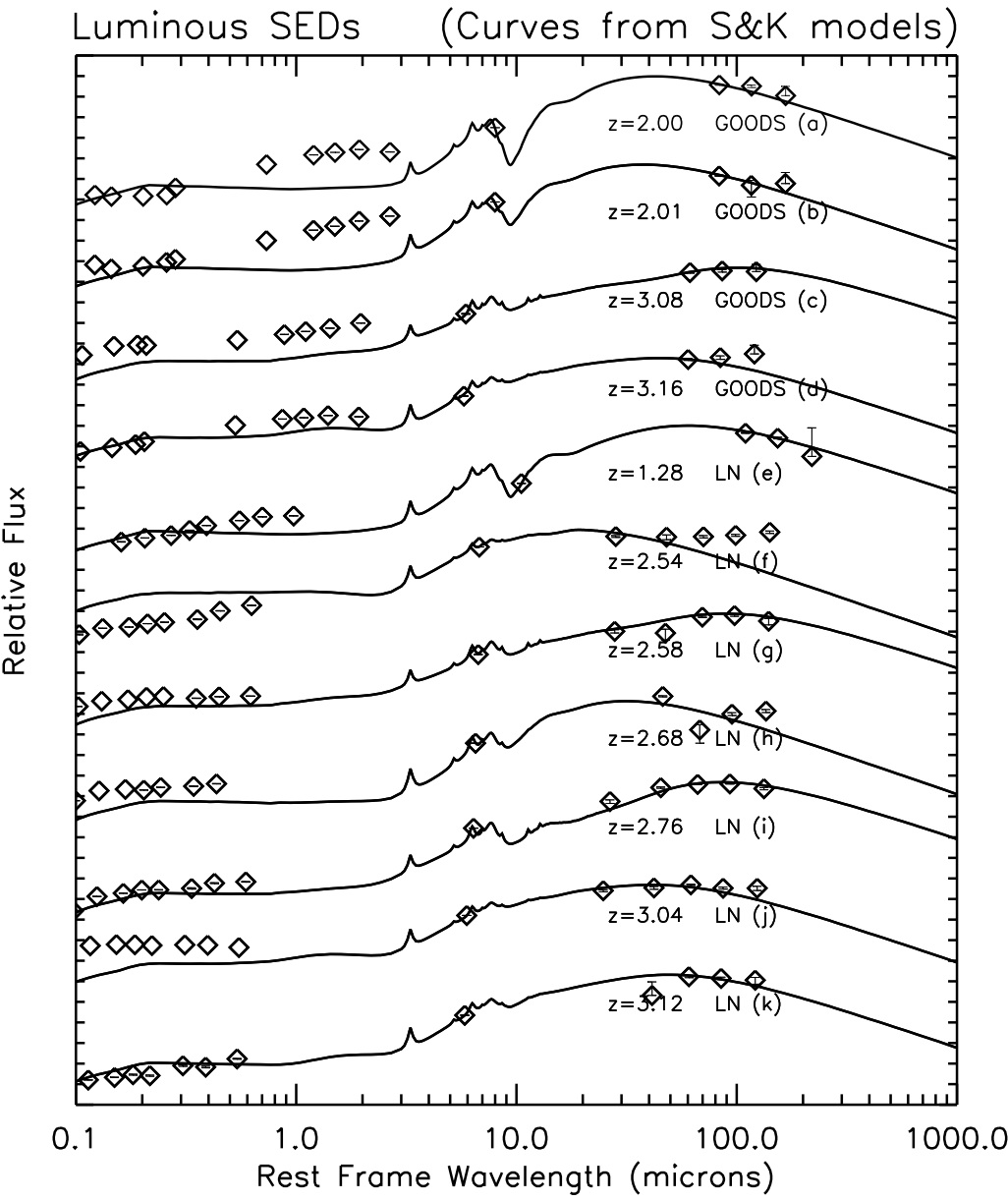}
	\caption{Spectral energy distributions of the most luminous sources in GOODS North (a) to (d) Lockman North (e) to (k).  As in Figures 2 and 3, we solely plot the upper error bars at 500 $\mu$m for sources (a) to (e), whose lower error bars are compatible with zero.  We similarly plot an upper error bar at 170 $\mu$m for LN (k).  Luminosities and positions for these sources are presented in Table \ref{tab:bigluminous}. Seven of these eleven sources are triply secure; two of these are also included in Figures 2 and 3.  A brief description of sources (c), (d), (f) and (i) is provided in Section 6.  Although some of the ultraluminous sources exhibit
significant AGN activity, we have applied S\&K model fits, as discussed in
Section 7. As in Figures 2 and 3 we have weighted the 24 $\mu$m data only
one quarter as heavily as the longer wavelengths.  Observational data shortward of 24 $\mu$m is plotted for reference but not used in fitting.}
	\label{fig:luminoussed}
\end{figure}

It is unlikely that three selection effects inherent in our observations
cast this conclusion into serious doubt:

(i) The first is that Figure \ref{fig:lumdist}, on which the conclusion is based, only
includes sources detected at all three SPIRE wavelengths. For highly
redshifted sources, the 100 $\mu$m infrared emission peak is redshifted into the
500 $\mu$m range, favoring the detection of galaxies at all three
wavelengths, including 500 $\mu$m.

(ii) However, compensating for this effect, lower redshift sources are more
readily detected by a factor inversely proportional to luminosity distance
squared.  Although these two effects partially cancel, high-luminosity
sources should be more readily detected at low than at high redshifts.

(iii) The XID catalogs search for SPIRE sources solely at locations where
Spitzer 24 $\mu$m sources exhibit flux densities $\geq$ 20 $\mu$Jy in the
GOODS-N field and $\geq$ 50 $\mu$Jy in LN. We may thus be missing sources at
redshifts at which poorly emitting spectral regions are redshifted to $24
\mu$m. At redshifts $z \sim 1.4$, for example, the 9.7 $\mu$m silicate
absorption dip shown by \citet{sp} to be prevalent in many ULIRGS is shifted
to 24 $\mu$m.  This may account for the striking absence of low-luminosity
sources, at $z \sim 1.4$, i.e. the lack of sources hugging the luminosity
distance curve at this redshift in Figure \ref{fig:lumdist}.

\section[]{Discussion}

Far-infrared surveys with Herschel need to take source confusion and source blending into account, particularly at the longest wavelengths, 350 and 500 $\mu$m. GOODS-N and LN are the two deepest surveys undertaken as part of the HerMES project to date. In these deep fields, crowding of sources presents especially serious problems.  In establishing a set of criteria that assess source blending, we have taken a preliminary step toward estimating the utility of the survey data for different purposes. This has proven itself useful in our analysis of the ultraluminous galaxies, some of which we described in Section 6 and whose characteristics are exhibited in Figure \ref{fig:luminoussed} and Table \ref{tab:bigluminous}. In view of the high infrared luminosities we find, it is particularly satisfying that seven of the eleven sources cited turn out to be triply secure, i.e. with high purities in all three SPIRE wave-bands, and that five of the sources also are observed by PACS where blending is not severe, particularly in the 100 $\mu$m waveband.  In
this context, we have placed no upper limit on acceptable values of $\Pi$,
which are especially high for GOODS(c) and LN(f), suggesting especially low
ambient source contributions at their locations.

In compiling the SEDs for GOODS-N and LN, we have elected to work with the S\&K models because they are based on a  limited set of well-defined physical parameters.  The models thus make predictions that our SEDs may be able to verify, refute, or extend.  S\&K do not specifically address the effects of adding an AGN component to a starburst model.  However, they do provide a starburst fit for NGC 6240 and propose that addition of a small AGN component could provide an improved fit.  Most starbursts generally also exhibit some AGN activity.  Perhaps because of this, the S\&K models appear to provide reasonable fits.  The major weakness of the S\&K models, as well as that of all others, tends to be the difficulty in accounting for the seemingly random relationship between the infrared and optical portions of the SEDs that is so apparent in Figures \ref{fig:stackedLNsed}, \ref{fig:stackedGOODSNsed}, and  \ref{fig:luminoussed}.

We investigated the relationship between flux-density ratios at optical and far-infrared wavelengths in high- and low-luminosity galaxies. Current theory suggests that starbursts involve stellar mass distributions obeying the Salpeter initial mass function \citep{z1}.
The drop in luminosities from the most massive O type stars with mass $\sim 120 M_\odot$ to the early B type stars at 10 $M_\odot$, can then be shown to be roughly in a ratio of 500:1, i.e. with a contrast considerably higher than that of the mass ratio, roughly 12:1.  The highest mass ranges will thus be depleted most rapidly,  ending their lives in supernova explosions in which at least some of the dust will be destroyed or expelled from the galaxy.  The most luminous galaxies found using our SEDs and their associated redshifts would thus be expected to be the very youngest as well as those most densely shrouded by dust, i.e. having the lowest fractional optical luminosities.  To test this hypothesis we restricted ourselves to galaxies with total bolometric luminosities, $L > 10^{12}L_\odot$, as these have long been considered likely starburst mergers, albeit with potential contributions from AGNs \citep{sm}.  In Table \ref{tab:fluxratio}, we compare the flux density ratios for the most and least luminous sources in three redshift bins.  It is evident that larger ratios correspond to more luminous sources at all redshifts but that these differences greatly diminish at lower redshifts.  This finding is both new and significant.  It indicates
evolutionary trends that may need to be incorporated into more advanced
models of starbursts designed to yield SEDs which not only mirror observed
ratios of optical to infrared emission, but also define a galaxy's place in its
evolutionary history.

\setcounter{table}{4}
\begin{table}
\caption{Ratios ($R\equiv F_{FIR}/F_{optical}$) of SPIRE flux densities (consistently measured at 250 $\mu$m to minimize source blending) to optical flux densities at a rest wavelength $\sim$3000 $\AA$, respectively measured at 4500 $\AA$, 6100 $\AA$, and 7600 $\AA$ for successively larger redshifts.  These three wavelengths were chosen to represent the visible spectra and avoid the Balmer jump at 3650 $\AA$. For each redshift bin we compare the flux density ratio for the $N/2$ most luminous galaxies ($R_{Bright}$) to that of the $N/2$ least luminous galaxies ($R_{Dim}$) where N is the total number of $L > 10^{12} L_\odot$ sources.}
\centering
\begin{tabular}{|l|l|l|l|l|l|l|l|l|l|}
\hline
 z & N  & $R_{Bright}$  & $R_{Dim}$ & $R_{Bright}/R_{Dim}$  \\ \hline
$0.95 < z < 1.05$ & 18 & 64500 & 72000 & 0.90 \\ 
$1.9 < z < 2.1$ & 18 & 62500 & 17600 & 3.6 \\ 
$2.4 < z < 2.8$ & 10 & 62300 & 5920 & 11 \\
\end{tabular}
\label{tab:fluxratio}
\end{table}

Figure \ref{fig:lumdist} provides a capsule history of galaxy evolution over cosmological  epochs for the sample included in our two deep surveys.    The shapes of these distributions are nearly identical in LN and GOODS-N, which motivated us to plot both in the same figure. A glance confirms that sources having the highest luminosities are found at highest redshifts, i.e. earliest epochs.  Luminosities higher than  $10^{13}L_{\odot}$ are generally observed at redshifts $z\sim 2 - 3.2$, the highest redshifts reached in our surveys.


\section{Conclusions}

Confusion, which can be troubling at 250 $\mu$m, becomes increasingly severe at 350 and 500 $\mu$m.  Yet the data at these longer wavelengths are  particularly important given how little is known about this spectral domain.   We believe that the triply-secure sources listed in Tables 2, 3, and 4 will be in demand for follow-on studies that X-ray astronomers, spectroscopists, and others may wish to undertake on sources known to be especially free of confusion.

With SPIRE photometry data in hand along with cross identifications at several shorter wavelengths, we have constructed SEDs for a handful of trustworthy sources in the GOODS-N and LN regions.  Many of these can be fit by starburst SED models, such as those created by \citet{s1},  which yield information on luminosity, dust mass, and size.   Figures 4 and 5 show a number of ultraluminous galaxies with $L_{IR} \sim 10^{13} L_{\odot}$.  Although these are extreme systems, they  do not appear to deviate from the general distribution at high redshift.  A major strength of the deep HerMES surveys is their ability to obtain reliable source luminosities and star-formation rates based on flux densities in the infrared and at auxiliary wavelengths as well as redshifts compiled in the XID catalogues.  


\section*{Acknowledgments}
This work is based in part on observations made with Herschel, a European
Space Agency Cornerstone Mission with significant participation by NASA.
Support for this work was provided by NASA through an award issued by
JPL/Caltech.

SPIRE has been developed by a consortium of institutes led by Cardiff University (UK) and including Univ. Lethbridge (Canada); NAOC (China); CEA, OAMP (France); IFSI, Univ. Padua (Italy); IAC (Spain); Stockholm Observatory (Sweden); Imperial College London, RAL, UCL-MSSL, UKATC, Univ. Sussex (UK); and Caltech/JPL, IPAC, Univ. Colorado (USA). This development has been supported by national funding agencies: CSA (Canada); NAOC (China); CEA, CNES, CNRS (France); ASI (Italy); MCINN (Spain); SNSB (Sweden); STFC (UK); and NASA (USA).

We thank the referee of this paper, Stephen J. Messenger, for his incisive comments and helpful recommendations.

\bsp


\clearpage
 
\setcounter{table}{1}
\begin{table*}\tiny
\caption{Data on LN Galaxies whose SEDs appear in Figure \ref{fig:stackedLNsed}.  We list the ID as established by \citet{i1}, right ascension and declination (J2000), photometric redshift $z>0.5$,  our purity indices $\Pi_{\lambda}$, total bolometric luminosity as estimated by an S\&K model, star formation rate based on the infrared luminosity relation \citep{k1}, and dust mass estimated by the S\&K model.   Along with these derived parameters, we list the model parameters: nuclear radius, visual extinction to center (A$_v$), and gas density within hotspots (n$_{hs}$).}  
\centering

\begin{tabular}{ccccccccccccccccc}
\hline
ID & RA & Dec & z & L$_{bolometric}$ ($L_{\odot}$) & SFR ($M_{\odot}$/yr) & M$_{dust}$ ($M_{\odot}$) & $\Pi_{250}$ & $\Pi_{350}$ & $\Pi_{500}$ & Radius (kpc) &  A$_v$ & Log$_{10}$(n$_{hs}\times$cm$^3$) & \\ \hline \hline

   LN (A) &  161.8365 &   59.1211 & 0.56 & 4.0$\times10^{11}$ &     66 & 1.6$\times10^{ 8}$ & 0.89 & 0.67 & 0.74 &   3.0 &  35.9 &  3  \\ \hline
   LN (B) &  161.5001 &   58.8732 & 0.60 & 7.9$\times10^{11}$ &    130 & 3.2$\times10^{ 8}$ & 0.74 & 0.62 & 0.88 &   3.0 &  72.0 &  4  \\ \hline
   LN (C) &  161.1277 &   59.1956 & 0.72 & 1.3$\times10^{12}$ &    210 & 3.2$\times10^{ 8}$ & 1.10 & 1.00 & 1.07 &   3.0 &  72.0 &  2  \\ \hline
   LN (D) &  161.3232 &   59.2086 & 0.82 & 1.3$\times10^{12}$ &    210 & 3.2$\times10^{ 8}$ & 0.88 & 0.80 & 0.98 &   3.0 &  72.0 &  4  \\ \hline
   LN (E) &  161.0530 &   59.0762 & 0.94 & 2.0$\times10^{12}$ &    330 & 1.6$\times10^{ 8}$ & 0.92 & 0.65 & 0.69 &   3.0 &  35.9 &  3  \\ \hline
   LN (F) &  161.3680 &   59.2242 & 0.96 & 2.5$\times10^{12}$ &    400 & 3.6$\times10^{ 8}$ & 0.96 & 0.97 & 0.56 &   9.0 &   9.0 &  4  \\ \hline
   LN (G) &  161.3429 &   59.2269 & 1.02 & 4.0$\times10^{12}$ &    650 & 1.6$\times10^{ 8}$ & 0.89 & 0.87 & 0.64 &   3.0 &  35.9 &  2  \\ \hline
   LN (H) &  161.4871 &   58.8886 & 1.06 & 2.5$\times10^{12}$ &    410 & 1.6$\times10^{ 8}$ & 0.76 & 0.82 & 0.93 &   3.0 &  35.9 &  2  \\ \hline
   LN (I) &  161.8669 &   58.8708 & 2.76 & 2.5$\times10^{13}$ &   4300 & 4.9$\times10^{ 9}$ & 0.91 & 0.96 & 0.77 &   9.0 & 120.0 &  4  \\ \hline

\end{tabular}
\label{tab:bigLN}

\end{table*}

\begin{table*}\tiny

\caption{Data on GOODS-N galaxies whose SEDs appear in Figure \ref{fig:stackedGOODSNsed}, listed as in Table \ref{tab:bigLN}.  The redshifts here are spectroscopic.} 

\centering
\begin{tabular}{cccccccccccccc}
\hline
ID & RA & Dec & z & L$_{bolometric}$ ($L_{\odot}$) & SFR ($M_{\odot}$/yr) & M$_{dust}$ ($M_{\odot}$) & $\Pi_{250}$ & $\Pi_{350}$ & $\Pi_{500}$ & Radius (kpc) & A$_v$ & Log$_{10}$(n$_{hs}\times$cm$^3$)  \\ \hline \hline
   
   GOODS (A) &  189.0274 &   62.1643 & 0.6380 & 5.0$\times10^{11}$ &     84 & 3.2$\times10^{ 8}$ & 0.83 & 0.84 & 0.44 &   3.0 &  72.0 &  4  \\ \hline
   GOODS (B) &  189.3938 &   62.2898 & 0.6402 & 2.0$\times10^{12}$ &    340 & 6.0$\times10^{ 7}$ & 0.98 & 0.64 & 0.90 &   1.0 & 119.0 &  2  \\ \hline
  GOODS (C) &  189.2979 &   62.1820 & 0.8549 & 1.3$\times10^{11}$ &     21 & 3.2$\times10^{ 8}$ & 1.00 & 0.81 & 0.74 &   3.0 &  72.0 &  4  \\ \hline
  GOODS (D) &  189.1403 &   62.1683 & 1.0160 & 1.6$\times10^{12}$ &    260 & 1.6$\times10^{ 8}$ & 0.99 & 0.75 & 0.54 &   3.0 &  35.9 &  3  \\ \hline
  GOODS (E) &  189.0633 &   62.1691 & 1.0270 & 1.3$\times10^{12}$ &    210 & 1.6$\times10^{ 8}$ & 0.92 & 0.52 & 0.45 &   3.0 &  35.9 &  2  \\ \hline
   GOODS (F) &  189.3171 &   62.3541 & 1.1440 & 1.0$\times10^{12}$ &    160 & 8.1$\times10^{ 7}$ & 0.73 & 0.68 & 0.68 &   3.0 &  17.9 &  2  \\ \hline
  GOODS (G) &  189.1438 &   62.2114 & 1.2242 & 2.5$\times10^{13}$ &   4300 & 6.0$\times10^{ 7}$ & 0.88 & 0.89 & 0.98 &   1.0 & 119.0 &  4  \\ \hline
  GOODS (H) &  189.2137 &   62.1810 & 1.2258 & 6.3$\times10^{11}$ &    100 & 1.6$\times10^{ 8}$ & 0.71 & 0.83 & 0.31 &   3.0 &  35.9 &  4  \\ \hline
   GOODS (I) &  189.2614 &   62.2338 & 1.2480 & 1.3$\times10^{12}$ &    200 & 3.6$\times10^{ 8}$ & 0.98 & 0.93 & 0.97 &   9.0 &   9.0 &  4  \\ \hline
  GOODS (J) &  189.2566 &   62.1962 & 1.7600 & 6.3$\times10^{12}$ &   1000 & 7.3$\times10^{ 8}$ & 1.02 & 1.04 & 1.03 &   9.0 &  18.0 &  4  \\ \hline
  GOODS (K) &  189.3036 &   62.1955 & 1.8150 & 2.0$\times10^{12}$ &    310 & 8.1$\times10^{ 7}$ & 1.08 & 0.86 & 0.50 &   3.0 &  17.9 &  3.4  \\ \hline        
  GOODS (L) &  189.0764 &   62.2640 & 2.0000 & 2.0$\times10^{13}$ &   3400 & 6.0$\times10^{ 7}$ & 0.82 & 0.71 & 0.33 &   1.0 & 119.0 &  2  \\ \hline

\end{tabular}
\label{tab:bigGOODSN}
\end{table*}

\begin{table*}\tiny

\caption{Data on luminous galaxies whose SEDs appear in Figure \ref{fig:luminoussed}, listed as in previous tables.  The GOODS sources have spectroscopic redshifts, as do LN (f) and (j).  The rest of the LN sources have photometric redshifts.  The sources GOODS (a) and LN (i) correspond to the sources GOODS (L) and LN (I) shown in the previous figures and tables.} 

\centering
\begin{tabular}{cccccccccccccc}
\hline
ID & RA & Dec & z & L$_{bolometric}$ ($L_{\odot}$) & SFR ($M_{\odot}$/yr) & M$_{dust}$ ($M_{\odot}$) & $\Pi_{250}$ & $\Pi_{350}$ & $\Pi_{500}$ & Radius (kpc) & A$_v$ & Log$_{10}$(n$_{hs}\times$cm$^3$)  \\ \hline \hline
   
      GOODS (a) &  189.0764 &   62.2640 & 2.0000 & 2.0$\times10^{13}$ &   3400 & 6.0$\times10^{ 7}$ & 0.82 & 0.71 & 0.33 &   1.0 & 119.0 & 2.0  \\ \hline
   GOODS (b) &  189.1483 &   62.2400 & 2.0050 & 2.5$\times10^{13}$ &   4200 & 3.6$\times10^{ 7}$ & 0.69 & 0.24 & 0.38 &   1.0 &  71.0 & 2.0  \\ \hline
  GOODS (c) &  188.9901 &   62.1734 & 3.0750 & 6.3$\times10^{12}$ &   1000 & 2.0$\times10^{ 9}$ & 1.61 & 1.22 & 0.84 &  15.0 &  18.0 & 4.0  \\ \hline
  GOODS (d) &  189.3096 &   62.2024 & 3.1569 & 1.3$\times10^{13}$ &   2000 & 4.0$\times10^{ 7}$ & 0.57 & 0.55 & 0.84 &   3.0 &   9.0 & 2.0  \\ \hline

   LN (e) &  161.7059 &   59.3247 & 1.28 & 1.3$\times10^{13}$ &   2100 & 3.2$\times10^{ 8}$ & 0.99 & 0.85 & 0.20 &   3.0 &  72.0 & 2.0  \\ \hline
   LN (f) &  161.0415 &   58.8735 & 2.28 & 2.5$\times10^{13}$ &   3400 & 4.1$\times10^{ 5}$ & 0.98 & 1.10 & 1.71 &   0.3 &   6.7 & 2.0  \\ \hline
   LN (g) &  161.5408 &   58.7950 & 2.58 & 1.3$\times10^{13}$ &   2000 & 2.0$\times10^{ 9}$ & 0.82 & 0.95 & 0.58 &  15.0 &  18.0 & 4.0  \\ \hline
   LN (h) &  160.9635 &   58.9555 & 2.68 & 2.5$\times10^{13}$ &   4300 & 8.8$\times10^{ 6}$ & 0.18 & 0.95 & 0.96 &   0.3 & 144.0 & 2.0  \\ \hline
   LN (i) &  161.8667 &   58.8704 & 2.76 & 2.5$\times10^{13}$ &   4300 & 4.9$\times10^{ 9}$ & 0.91 & 0.96 & 0.77 &   9.0 & 120.0 & 4.0  \\ \hline
   LN (j) &  161.5541 &   58.7886 & 2.96 & 2.0$\times10^{13}$ &   3100 & 4.0$\times10^{ 7}$ & 0.76 & 0.61 & 0.60 &   3.0 &   9.0 & 3.0  \\ \hline
   LN (k) &  161.8259 &   59.2771 & 3.12 & 1.3$\times10^{13}$ &   2000 & 8.0$\times10^{ 7}$ & 0.98 & 0.70 & 0.54 &   3.0 &  17.8 & 2.0  \\ \hline

\end{tabular}
\label{tab:bigluminous}
\end{table*}


\label{lastpage}


\begin{thebibliography}{99}
\bibitem[\protect\citeauthoryear{Barger et al.}{2008}]{b1} Barger, A.J., et al., 2008,  ApJ, 689, 687
\bibitem[\protect\citeauthoryear{B\'{e}thermin et al.}{2010}]{b2} B\'{e}thermin, M., et al., 2010,
A\&A,516, 43
\bibitem[\protect\citeauthoryear{Buat et al.}{2010}]{buat} Buat V., et al., 2010,
MNRAS, this issue
\bibitem[\protect\citeauthoryear{Fixsen et al.}{1998}]{f1} Fixsen, D. J., et al., 1998,
ApJ, 508, 123
\bibitem[\protect\citeauthoryear{Griffin et al.}{2010}]{spire} Griffin, et al., 2010, A\&A 518, L3
\bibitem[\protect\citeauthoryear{Kennicutt}{1998}]{k1} Kennicutt, R., 1998,
Ann. Rev. A\&A, 36, 189
\bibitem[\protect\citeauthoryear{Law et al.}{2007}]{l1} Law, D. R., et al., 2007, ApJ 656, 1
\bibitem[\protect\citeauthoryear{Marsden et al.}{2009}]{m1} Marsden, G., et al., 2009,
ApJ, 707, 1729
\bibitem[\protect\citeauthoryear{Morrison et al.}{2010}]{m2}  Morrison, G. E., et al., 2010, ApJS 188,178
\bibitem[\protect\citeauthoryear{Oliver et al.}{2010}]{hermes} Oliver, et al., 2010, A\&A 518, L2
\bibitem[\protect\citeauthoryear{Oliver et al.}{2010a}]{o2} Oliver, et al., 2010a,
in preparation
\bibitem[\protect\citeauthoryear{Owen \& Morrison}{2008}]{o1} Owen, F. N. \& Morrison, G. E., 2008, AJ 136,1889
\bibitem[\protect\citeauthoryear{Pilbratt et al.}{2010}]{herschel} Pilbratt, et al., 2010, A\&A 518, L1
\bibitem[\protect\citeauthoryear{Poglitsch et al.}{2010}]{pacs} Poglitsch, et al., 2010, A\&A 518, L2
\bibitem[\protect\citeauthoryear{Polletta et al.}{2006}]{p1} Polletta, M., et al., 2006,
ESASP 604, 807
\bibitem[\protect\citeauthoryear{Roseboom et al.}{2010}]{i1} Roseboom I., et al., 2010,
MNRAS, this issue
\bibitem[\protect\citeauthoryear{Sanders \& Mirabel}{1996}]{sm} Sanders, D. B. \& Mirabel, I. F., 1996, ARA\&A 1996, 34, 749
\bibitem[\protect\citeauthoryear{Bernhard Schulz}{private communication}]{sc1} Schulz B., 2010,
private communication
\bibitem[\protect\citeauthoryear{Siebenmorgen \& Kr\"{u}gel}{2007}]{s1} Siebenmorgen R., Kr\"{u}gel E., 2007,
A\&A, 461, 445
\bibitem[\protect\citeauthoryear{Smith et al.}{2010}]{s2} Smith A. J., et al., 2010,
MNRAS, this issue
\bibitem[\protect\citeauthoryear{Spoon et al.}{2007}]{sp} Spoon, H. W. W.,
et al., 2007, ApJ 654, L49
\bibitem[\protect\citeauthoryear{Strazzullo et al.}{2010}]{st} Strazzullo, V.,
et al., 2010, ApJ 714, 1305
\bibitem[\protect\citeauthoryear{Takeuchi, T. T. \& Ishii, T. T.}{2004}]{tt}
Takeuchi, T. T. \&Ishii, T. T., 2004, ApJ 604, 40 
\bibitem[\protect\citeauthoryear{Trouille et al.}{2008}]{t1} Trouille, L., et al., 2008, ApJS 179, 1
\bibitem[\protect\citeauthoryear{Zinnecker \& Yorke}{2007}]{z1}
Zinnecker, H., \& Yorke, H., 2007 Ann. Rev. A\&A, 45, 481
\end{thebibliography}
\end{document}